\documentclass[
  aps,
  prd,
  preprint,
  nofootinbib
]{revtex4}

\usepackage{amsmath,amssymb}
\usepackage{graphicx}
\usepackage{slashed}

\DeclareMathOperator{\tr}{tr}

\begin{document}

\preprint{Brown/HET-1581}

\title{Higher point MHV amplitudes in $\mathcal{N}=4$ Supersymmetric Yang-Mills Theory}

\author{C.~Vergu}
\email{Cristian_Vergu@brown.edu}
\affiliation{Physics Department, Brown University, Providence, RI 02912, USA}

\begin{abstract}
We compute the even part of the two-loop seven-point planar MHV amplitude in $\mathcal{N}=4$ supersymmetric Yang-Mills theory.  We find that the even part is expressed in terms of conformal integrals with simple rational coefficients.  We also compute the even part of two all-$n$ cuts.  An important feature of the result is that no hexagon (or higher polygon) loops appear among the integrals detected by the cuts we computed.

We also present a ``leg addition rule,'' which allows us to express some integral coefficients in the $n+1$-point MHV amplitude in terms of the integral coefficients of the $n$-point MHV amplitude.
\end{abstract}


\maketitle

\section{Introduction}

There are strong indications that one can hope to understand the maximally supersymmetric $\mathcal{N}=4$ gauge theory, at least in the 't Hooft limit~\cite{'tHooft:1973jz}.  This hope is motivated by Maldacena's AdS/CFT duality between the $\mathcal{N}=4$ gauge theory and type IIB string theory on $AdS_5 \times \mathbb{S}^5$ background (see refs.~\cite{Maldacena:1997re, Gubser:1998bc, Witten:1998qj}).

While there are many instances of the AdS/CFT duality, the one involving the $\mathcal{N}=4$ theory is special in that it has the maximal symmetry possible for a four-dimensional field theory.  Moreover, integrability has been shown to play an important role on both sides of this duality (see refs.~\cite{Minahan:2002ve, Bena:2003wd}).

In ref.~\cite{Witten:2003nn} Witten showed that the $\mathcal{N}=4$ theory has surprising properties when transformed to Penrose's twistor space and he also proposed a dual string theory in twistor space.  This duality, despite of being of weak--weak type is still poorly understood.

Work on $\mathcal{N}=4$ scattering amplitudes has concentrated in two main directions: a better understanding of tree-level amplitudes and a better understanding of loop-level amplitudes and all-loop ansatze.

A comprehensive understanding of the tree-level amplitudes has been achieved so far (see refs.~\cite{Cachazo:2004kj, Britto:2005fq, Drummond:2008vq, Drummond:2008cr}), but from recent developments (see refs.~\cite{ArkaniHamed:2008gz, ArkaniHamed:2009si, Mason:2009sa, Drummond:2009fd}) it is clear that much remains to be uncovered.

On the higher-loop front surprising success has been achieved as well.  This has culminated with the formulation of the ABDK~\cite{Anastasiou:2003kj} and BDS~\cite{Bern:2005iz} ansatze and, more recently, with the conjecture of a dual conformal supersymmetry by DHKS~\cite{Drummond:2008vq}.

For a while, the evidence in favor of the all-loop BDS ansatz only came from weak 't Hooft coupling computations (see refs.~\cite{Anastasiou:2003kj, Bern:2005iz, Bern:2006vw, Cachazo:2006tj}).  This changed after a striking paper~\cite{Alday:2007hr} by Alday and Maldacena, where they used the AdS/CFT correspondence to compute the scattering amplitudes at strong 't Hooft coupling.  They found that the BDS ansatz also holds at strong coupling for four-point amplitudes.  This paper also hinted that there might be a very strong link between scattering amplitudes and Wilson loops in $\mathcal{N}=4$ gauge theory.  This Wilson-loops--scattering amplitudes correspondence was soon confirmed to hold at weak coupling also (see refs.~\cite{Drummond:2007aua, Brandhuber:2007yx, Drummond:2007cf}).

In ref.~\cite{Drummond:2007au}, using a conformal Ward identity for Wilson loops, it was shown that, assuming that the Wilson-loops--scattering amplitudes correspondence holds, the BDS ansatz also holds at five points but may fail starting at six points.  In ref.~\cite{Alday:2007he}, Alday and Maldacena also gave some arguments that the BDS ansatz has to fail for a large number of external legs.

In refs.~\cite{Drummond:2007bm, Drummond:2008aq, Bern:2008ap} it was demonstrated by explicit perturbative computations that the BDS ansatz fails at weak coupling starting at six points but, miraculously, the Wilson-loops--scattering amplitudes correspondence survives.  Later, it was shown in ref.~\cite{Cachazo:2008hp} that the odd part of the two-loop six-point amplitude satisfies the BDS ansatz, provided there are no odd ``$\mu$ integrals.''  In ref.~\cite{unpublished:2008} it was shown that such ``$\mu$ integrals'' do not appear in the odd part of the six-point MHV amplitude.

More recently, the dual conformal supersymmetry was understood from a string theory perspective as a consequence of the fact that the $AdS_5 \times \mathbb{S}^5$ background is mapped to itself by some fermionic $T$-duality transformations (see ref.~\cite{Berkovits:2008ic}), combined with the non-compact $T$-dualities in the space-time directions, as described in ref.~\cite{Alday:2007hr}.  In refs.~\cite{Ricci:2007eq, Berkovits:2008ic, Beisert:2008iq} the action of the fermionic $T$-duality on the non-local charges of the integrable worldsheet action was studied.

Our motivation for computing these higher-point MHV amplitudes is justified in part by the desire to compare to the higher-point Wilson loop results presented in ref.~\cite{Anastasiou:2009kn}.  So far, the link between the integrals used to represent the amplitudes and the integrals used to represent the Wilson loops has not been made explicit.  We hope that the information presented in this paper, together with the results of ref.~\cite{Anastasiou:2009kn} can help to further our understanding of this important question.

This is a rapidly evolving subject, but we recommend the interested reader a set of recent reviews~\cite{Dixon:2008tu, Alday:2008yw, Bern:2007dw} for more details than we could give in this short introduction.

Let us finally note that the scattering amplitudes in the three-dimensional $\mathcal{N}=6$ supersymmetric ABJM theory~\cite{Aharony:2008ug} were studied in ref.~\cite{Agarwal:2008pu}.

The plan of the paper is as follows: in Sec.~\ref{sec:seven-point} we compute the seven-point two loop MHV amplitude, in Sec.~\ref{sec:all-n} we compute two all-$n$ cuts, in Sec.~\ref{sec:leg-add} we conjecture a new rule allowing us to obtain $n+1$-point amplitudes starting with $n$-point amplitudes.  We end with conclusions and discussion.

\section{The seven-point MHV amplitude}
\label{sec:seven-point}

\subsection{Computational procedure}

In this paper we will only discuss planar $\mathcal{N}=4$ scattering amplitudes.  The non-planar amplitudes, while interesting in themselves and for the $\mathcal{N}=8$ supergravity computations, are much less understood.\footnote{However, some progress has been made in understanding the non-planar amplitudes.  Most notably, ref.~\cite{Bern:2008qj} has discovered a Jacobi-like identity, which helps to relate the non-planar integral coefficients to the the planar integral coefficients which are easier to compute.}  The one-loop $n$-point MHV scattering amplitudes were computed in ref.~\cite{Bern:1994zx}.  See also ref.~\cite{Bern:1994cg} for a six-point one-loop NMHV computation and ref.~\cite{Drummond:2008bq} for a $n$-point NMHV computation using a $\mathcal{N}=4$ chiral superspace notation.

\begin{figure}
  \centering
  \includegraphics[width=.8\textwidth]{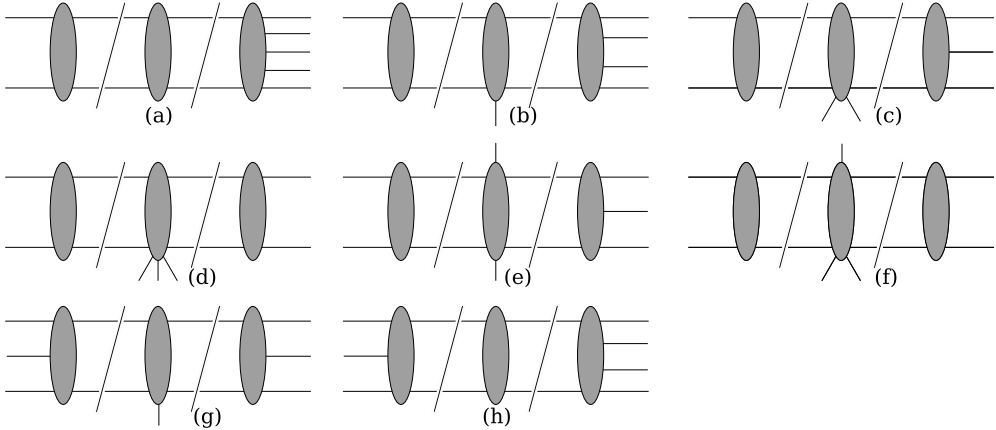}
  \caption{The unitarity cuts used to compute the seven-point MHV amplitude.}
  \label{fig:cuts-7pt}
\end{figure}

We computed the integral coefficients by using the unitarity method~\cite{Bern:1994zx, Bern:1997sc} and the iterated two-particle cuts in Fig.~\ref{fig:cuts-7pt}.  More precisely, we have computed only the four-dimensional cuts.

The four-dimensional cut computation is known to drop some contributions coming from integrals whose integrand vanishes in the limit where the dimensional regulator is removed, $\epsilon \to 0$, but whose integral can still give finite or divergent contributions in that limit.  An example of this phenomenon is provided by the six-point two-loop computation from ref.~\cite{Bern:2008ap}.

However, as the computation in ref.~\cite{Bern:2008ap} showed, the contribution of these so-called two-loop ``$\mu$ integrals'' cancels against the product of the one-loop $\mu$ integrals with the one-loop integrals which are detected by the four-dimensional cuts.  Therefore, for the purposes of computing the two-loop correction to the BDS ansatz (or the remainder function as it has been called in ref.~\cite{Bern:2008ap}), these $\mu$ integrals can be dropped.  Another way of phrasing this is that at six points, at least through two-loops, the $\mu$ integrals don't contribute to the finite or divergent part of the logarithm of the amplitude.  We take this as a strong hint that the same cancellation will happen at seven and higher points, at least for MHV amplitudes.

In order to compute the $\mu$ integrals, one would have to perform a $D$-dimensional cut computation, which is significantly harder than the four-dimensional cut computation.

Starting at five points, the scattering amplitudes contain both even and odd parts.  While for the even part only dual conformal integrals with simple rational coefficients seem to appear (see refs.~\cite{Drummond:2006rz, Drummond:2007aua} for a discussion), no such simple prescription is known for the odd parts.  In fact, the odd part is known analytically only for five-point amplitudes (see refs.~\cite{Bern:2006vw, Cachazo:2008vp} for the two-loop and ref.~\cite{Spradlin:2008uu} for the three-loop odd parts).

In ref.~\cite{Cachazo:2008hp} the odd part of the six-point two-loop amplitude was computed by using the leading singularity method of Cachazo (see ref.~\cite{Cachazo:2008vp} and also refs.~\cite{Cachazo:2008dx, Buchbinder:2005wp}).

Summarizing, we compute the even part of the seven-point MHV scattering amplitude by starting with a conformal ansatz for the integrals detected by a cut and matching it to the product of tree amplitudes.  It is easiest to find the coefficients numerically and this can be achieved by generating some random kinematics, subject to momentum conservation and on-shell conditions, and using that kinematics to evaluate both the conformal ansatz and the product of tree amplitudes.  By imposing the condition that these two contributions are identical, and by generating a sufficient number of equations, we can find all the undetermined coefficients by solving a linear system of equations.

We choose to work with the set of cuts in Fig.~\ref{fig:cuts-7pt} for several reasons.  First, these cuts can be computed using only MHV tree amplitudes.  Second, under the assumption that there are no integrals containing triangle loops, these cuts are sufficient to determine all the integrals contributing to the amplitude (in fact, the results obtained by computing all the cuts in Fig.~\ref{fig:cuts-7pt} contain a great deal of redundant information, which can be used as a cross-check on the computation).

\begin{figure}
  \centering
  \includegraphics{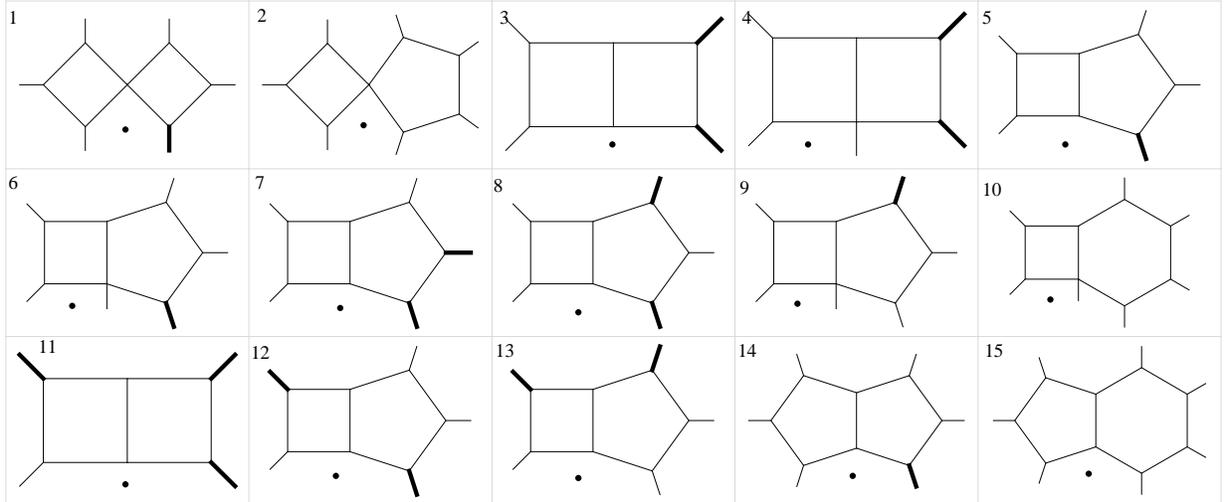}
  \caption{The integral topologies which can be made conformal, but appear with zero coefficient in the result for the seven-point MHV amplitude.  Sometimes two or more external massless legs are attached at the same point and their sum is generically a massive momentum.  These massive momenta are denoted by thick solid lines and the massless legs are denoted by thin solid lines.  The dot marks the position of the dual variable $x_1$, while the rest of the dual variables are ordered clockwise.}
  \label{fig:zero-7pt}
\end{figure}

\begin{figure}
  \centering
  \includegraphics{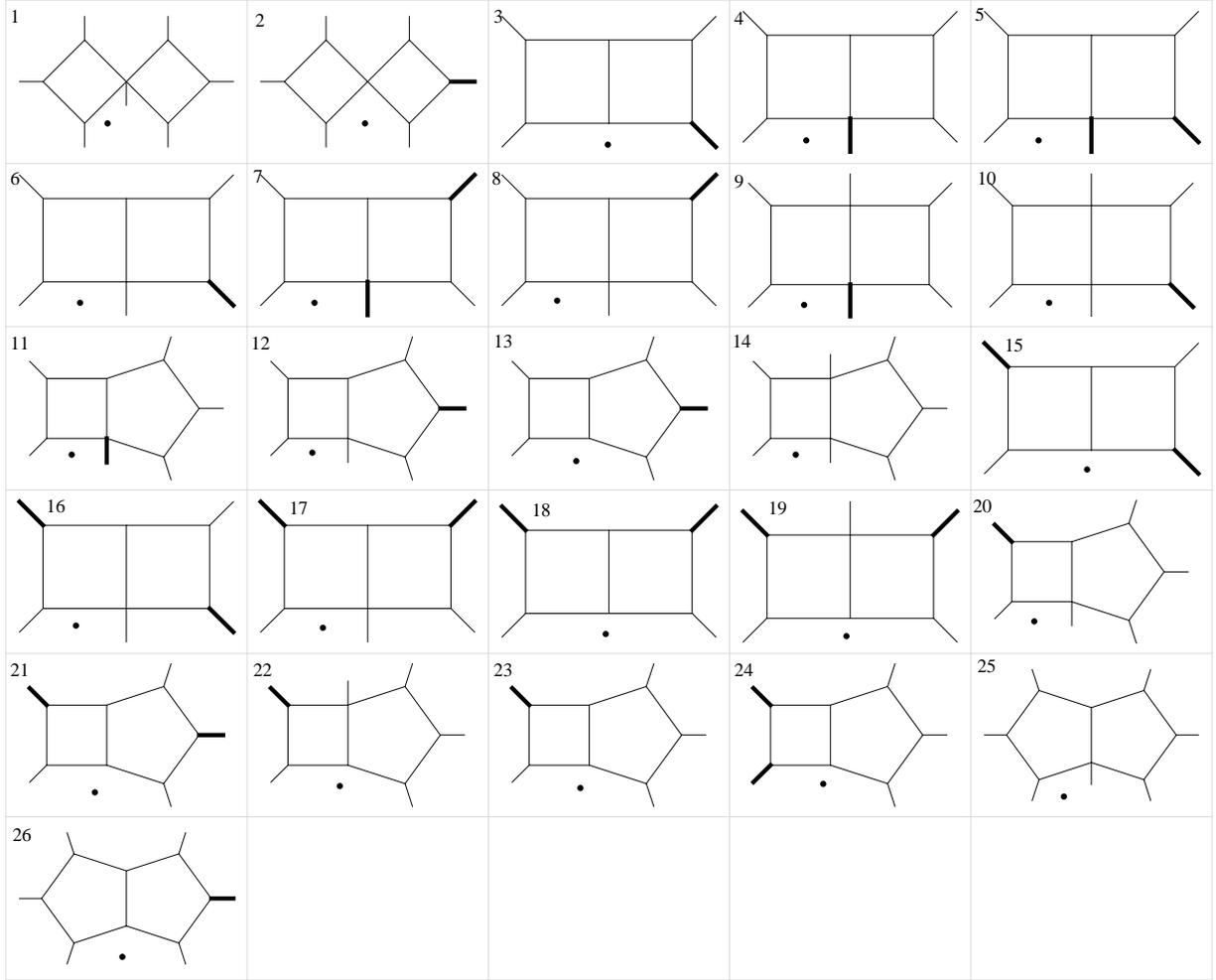}
  \caption{The integral topologies which can be made conformal and appear with nonzero coefficient in the result for the MHV amplitude.  The notations for the integrals are the same as those in Fig.~\ref{fig:zero-7pt}.  For the integrals $15$ and $18$ the notation is ambiguous because one can distribute the external legs in several ways to form massive momenta.  In those cases we take the convention that the massive leg which appears first when going clockwise starting at $x_1$ is a sum of two massless momenta, while the next massive leg is a sum of three massless momenta.  The dual variable of the left (right) loop is $x_p$ ($x_q$).}
  \label{fig:nonzero-7pt}
\end{figure}

\subsection{Results for the coefficients}

In table~\ref{tab:no-coefficients} we present some statistics about the number of coefficients.  It is obvious by inspection that a large number of coefficients vanish.  Indeed, for the topologies shown in Fig.~\ref{fig:zero-7pt} all the coefficients vanish.

\begin{table}
  \centering
  \begin{tabular}{|r||c|c|c|c|c|c|c|c|c|}
    \hline
    cut & a & b & c & d & e & f & g & h\\
    \hline
    coeff $= 0$ & 6 & 64 & 107 & 144 & 106 & 188 & 74 & 52\\
    \hline
    coeff $ \neq 0$ & 3 & 28 & 60 & 37 & 149 & 169 & 79 & 23\\
    \hline
  \end{tabular}
  \caption{The number of conformal coefficients detected by each of the cuts in Fig.~\ref{fig:cuts-7pt}.  We show the number of coefficients which are zero and the number of non-zero coefficients separately.  The counting above does not take into account the coefficients related by symmetries.}
  \label{tab:no-coefficients}
\end{table}

The numerators for the integral topologies in the notation of Fig.~\ref{fig:nonzero-7pt} are\footnote{Note that the coefficients below are presented in a different form from that used in e.g. ref.~\cite{Bern:2008ap}.  In that reference, the integrals included the parts of the numerator which were dependent on the loop momenta.  Because of this the coefficients were simpler but the list of integrals was longer.  We decided to keep the list of integrals in Fig.~\ref{fig:nonzero-7pt} as short as possible and, as a result, the integral ``coefficients'' depend on the loop momenta and contain terms which are related by permutations which leave the underlying integral topology invariant.}

{\allowdisplaybreaks 
\begin{align}
  c_1 = & \begin{aligned}[t]&\left(\left(x_{2 5}^2 x_{3 7}^2-x_{2 7}^2 x_{3 5}^2\right) x_{4 6}^2+\left(x_{2 6}^2 x_{3 5}^2-x_{2 5}^2 x_{3 6}^2\right) x_{47}^2\right) x_{1 4}^2+\\&x_{2 4}^2 \left(\left(x_{3 6}^2 x_{4 7}^2-x_{3 7}^2 x_{4 6}^2\right) x_{1 5}^2-x_{1 6}^2 x_{3 5}^2 x_{4 7}^2+x_{1 3}^2 x_{4 6}^2 x_{5 7}^2\right), \end{aligned}\\
  c_2 = & x_{1 4}^2 \left(\left(x_{2 5}^2 x_{4 7}^2-x_{2 4}^2 x_{5 7}^2\right) x_{1 3}^2+x_{1 5}^2 x_{2 4}^2 x_{3 7}^2+x_{1 4}^2 \left(x_{2 7}^2 x_{3 5}^2-x_{2 5}^2 x_{3 7}^2\right)\right),\\
  c_3 = & 2 x_{1 3}^4 x_{2 4}^2,\\
  c_4 = & x_{1 3}^2 x_{2 4}^2 x_{3 5}^2,\\
  c_5 = & -x_{1 3}^2 x_{2 4}^2 x_{3 6}^2,\\
  c_6 = & -x_{1 3}^2 x_{2 4}^2 x_{3 7}^2,\\
  c_7 = & x_{1 3}^2 \left(x_{2 5}^2 x_{3 6}^2-x_{2 6}^2 x_{3 5}^2\right),\\
  c_8 = & x_{1 3}^2 \left(x_{2 6}^2 x_{3 7}^2-2 x_{2 7}^2 x_{3 6}^2\right),\\
  c_9 = & \left(x_{2 6}^2 x_{3 5}^2-x_{2 5}^2 x_{3 6}^2\right) x_{1 4}^2-x_{1 6}^2 x_{2 4}^2 x_{3 5}^2+x_{1 5}^2 x_{2 4}^2 x_{3 6}^2-x_{1 3}^2 x_{2 5}^2 x_{4 6}^2,\\
  c_{10} = & \left(x_{2 4}^2 x_{5 7}^2-2 x_{2 5}^2 x_{4 7}^2\right) x_{1 3}^2+x_{1 4}^2 \left(x_{2 5}^2 x_{3 7}^2-x_{2 7}^2 x_{3 5}^2\right),\\
  c_{11} = & x_{1 3}^2 x_{2 q}^2 x_{3 5}^2 x_{4 6}^2,\\
  c_{12} = & x_{1 q}^2 x_{2 4}^2 x_{3 6}^2 x_{3 7}^2-x_{2 q}^2 \left(\left(x_{3 7}^2 x_{4 6}^2-2 x_{3 6}^2 x_{4 7}^2\right) x_{1 3}^2+x_{1 4}^2 x_{3 6}^2 x_{3 7}^2\right),\\
  c_{13} = & -2 x_{1 3}^2 x_{2 q}^2 \left(x_{1 3}^2 x_{4 7}^2-x_{1 4}^2 x_{3 7}^2\right),\\
  c_{14} = & \begin{aligned}[t]&\left(\left(2 x_{3 6}^2 x_{4 7}^2-x_{3 7}^2 x_{4 6}^2\right) x_{2 5}^2+x_{2 7}^2 x_{3 5}^2 x_{4 6}^2-2 x_{2 6}^2 x_{3 5}^2 x_{4 7}^2-x_{2 4}^2 x_{3 6}^2 x_{5 7}^2\right) x_{1 q}^2+\\&x_{2 q}^2 \left(\left(x_{3 7}^2 x_{4 6}^2-2 x_{3 6}^2 x_{4 7}^2\right) x_{1 5}^2+2 x_{1 6}^2 x_{3 5}^2 x_{4 7}^2+x_{1 4}^2 x_{3 6}^2 x_{5 7}^2\right)+\\&x_{3 q}^2 \left(\left(2 x_{2 6}^2 x_{4 7}^2-x_{2 7}^2 x_{4 6}^2\right) x_{1 5}^2-x_{1 4}^2 x_{2 6}^2 x_{5 7}^2+x_{1 6}^2 \left(x_{2 4}^2 x_{5 7}^2-2 x_{2 5}^2 x_{4 7}^2\right)\right), \end{aligned}\\
  c_{15} = & x_{1 4}^4 x_{2 5}^2-x_{1 4}^2 x_{1 5}^2 x_{2 4}^2,\\
  c_{16} = & x_{2 4}^2 \left(x_{1 4}^2 x_{5 7}^2-x_{1 5}^2 x_{4 7}^2\right),\\
  c_{17} = & x_{1 4}^2 \left(x_{2 6}^2 x_{4 7}^2-x_{2 7}^2 x_{4 6}^2\right)-x_{1 6}^2 x_{2 4}^2 x_{4 7}^2,\\
  c_{18} = & x_{1 4}^4 x_{2 7}^2,\\
  c_{19} = & -x_{1 4}^2 x_{1 5}^2 x_{2 7}^2,\\
  c_{20} = & \left(x_{1 4}^2 x_{2 q}^2-x_{1 q}^2 x_{2 4}^2\right) x_{4 6}^2 x_{5 7}^2,\\
  c_{21} = & x_{1 4}^2 x_{2 q}^2 \left(x_{1 5}^2 x_{4 7}^2-x_{1 4}^2 x_{5 7}^2\right),\\
  c_{22} = & x_{1 6}^2 \left(\left(x_{2 7}^2 x_{4 q}^2-x_{2 q}^2 x_{4 7}^2\right) x_{1 5}^2+x_{1 4}^2 x_{2 q}^2 x_{5 7}^2\right),\\
  c_{23} = & x_{1 5}^2 x_{1 6}^2 x_{2 q}^2 x_{5 7}^2,\\
  c_{24} = & \begin{aligned}[t] &x_{1 q}^2 x_{2 4}^2 \left(x_{3 7}^2 x_{4 6}^2-x_{3 6}^2 x_{4 7}^2\right) x_{5 p}^2+x_{1 4}^2 x_{2 q}^2 \left(x_{3 6}^2 x_{4 7}^2-x_{3 7}^2 x_{4 6}^2\right) x_{5 p}^2+\\&x_{3 q}^2 \left(\left(x_{2 7}^2 x_{4 6}^2-x_{2 6}^2 x_{4 7}^2\right) x_{1 4}^2+x_{1 6}^2 x_{2 4}^2 x_{4 7}^2\right) x_{5 p}^2+x_{1 q}^2 x_{2 4}^2 x_{3 5}^2 x_{4 7}^2 x_{6 p}^2-\\&x_{1 4}^2 x_{2 q}^2 x_{3 5}^2 x_{4 7}^2 x_{6 p}^2+\left(x_{1 4}^2 x_{2 5}^2-x_{1 5}^2 x_{2 4}^2\right) x_{3 q}^2 x_{4 7}^2 x_{6 p}^2-x_{1 q}^2 x_{2 4}^2 x_{3 5}^2 x_{4 6}^2 x_{7 p}^2+\\&x_{1 4}^2 x_{2 q}^2 x_{3 5}^2 x_{4 6}^2 x_{7 p}^2+\left(x_{1 5}^2 x_{2 4}^2-x_{1 4}^2 x_{2 5}^2\right) x_{3 q}^2 x_{4 6}^2 x_{7 p}^2,\end{aligned}\\
  c_{25} = & \begin{aligned}[t] &x_{1 q}^2 x_{2 4}^2 \left(x_{3 7}^2 x_{4 6}^2-x_{3 6}^2 x_{4 7}^2\right) x_{5 p}^2+x_{1 4}^2 x_{2 q}^2 \left(x_{3 6}^2 x_{4 7}^2-x_{3 7}^2 x_{4 6}^2\right) x_{5 p}^2+\\&x_{3 q}^2 \left(\left(x_{2 7}^2 x_{4 6}^2-x_{2 6}^2 x_{4 7}^2\right) x_{1 4}^2+x_{1 6}^2 x_{2 4}^2 x_{4 7}^2\right) x_{5 p}^2+x_{1 q}^2 x_{2 4}^2 x_{3 5}^2 x_{4 7}^2 x_{6 p}^2-\\&x_{1 4}^2 x_{2 q}^2 x_{3 5}^2 x_{4 7}^2 x_{6 p}^2+\left(x_{1 4}^2 x_{2 5}^2-x_{1 5}^2 x_{2 4}^2\right) x_{3 q}^2 x_{4 7}^2 x_{6 p}^2-x_{1 q}^2 x_{2 4}^2 x_{3 5}^2 x_{4 6}^2 x_{7 p}^2+\\&x_{1 4}^2 x_{2 q}^2 x_{3 5}^2 x_{4 6}^2 x_{7 p}^2+\left(x_{1 5}^2 x_{2 4}^2-x_{1 4}^2 x_{2 5}^2\right) x_{3 q}^2 x_{4 6}^2 x_{7 p}^2,\end{aligned}\\
  c_{26} = & \begin{aligned}[t] &-x_{2 7}^2 x_{3 q}^2 x_{5 p}^2 x_{1 4}^4-x_{2 q}^2 x_{3 5}^2 x_{7 p}^2 x_{1 4}^4+x_{2 q}^2 \left(x_{1 4}^4 x_{3 7}^2-x_{1 3}^2 x_{1 4}^2 x_{4 7}^2\right) x_{5 p}^2+\\&\left(x_{1 4}^4 x_{2 5}^2-x_{1 4}^2 x_{1 5}^2 x_{2 4}^2\right) x_{3 q}^2 x_{7 p}^2.\end{aligned}
\end{align}
}

Notice that the two-loop seven-point even coefficients are determined uniquely.  This reinforces the belief that the conformal basis of integrals is a ``good'' basis in the sense that there are no ambiguities related to reduction identities, like the ones that were found for the odd part of the two-loop six point amplitude (see ref.~\cite{Cachazo:2008hp}).

In the past, a powerful test that we obtained the complete result was by using the infrared consistency conditions.  However, here we have not evaluated the integrals so we can not use this test.

A modest check on the computation is provided by the agreement of the final answer with the rung-rule.  We have also checked the coefficients by computing the maximal cuts for the eight-propagator two-loop topologies (note that for the eight-propagator two-loop topologies the leading singularity is similar to the maximal cuts technique).

\section{Some all-$n$ cuts}
\label{sec:all-n}

\begin{figure}
  \centering
  \includegraphics[width=\textwidth]{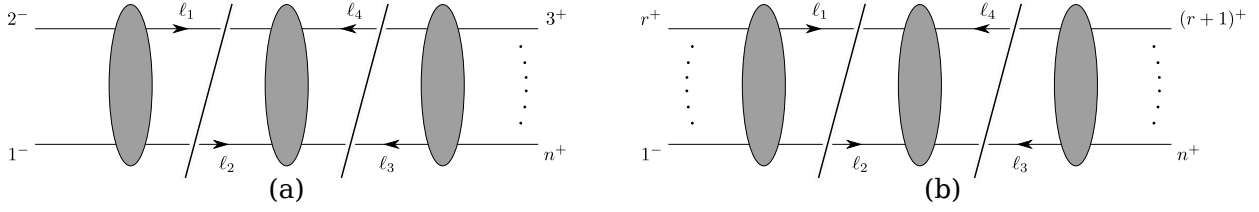}
  \caption{The two-loop $n$-points cuts used to constrain the $n$-point MHV amplitude.  For the cut (b) we take $r$ such that $2 < r < n-2$.}
  \label{fig:npt-cut}
\end{figure}

Here we compute the two-loop cuts in Fig.~\ref{fig:npt-cut} for an arbitrary number $n$ of external particles.

Let us start with the cut (a) in Fig.~\ref{fig:npt-cut} with the helicity assignment $(1^-, 2^-, 3^+, \dotsc, n^+)$.  With this helicity assignment the sums over internal states in the cuts have only one non-vanishing term.

The product of tree amplitudes corresponding to the cut (a) in Fig.~\ref{fig:npt-cut} is
\begin{equation}
  \frac {i \langle 1 2\rangle^3}{\langle 2 l_1\rangle \langle l_1 l_2\rangle \langle l_2 1\rangle} \frac {i \langle l_2 l_1\rangle^3}{\langle l_1 l_4\rangle \langle l_4 l_3\rangle \langle l_3 l_2\rangle} \frac {i \langle l_3 l_4\rangle^3}{\langle l_4 3\rangle \cdots \langle n l_3\rangle}.
\end{equation}

After dividing by the tree amplitude and rationalizing the denominators the expression above can be written as
\begin{equation}
  \label{eq:trace-cut-a}
  s_{1 2} \frac {\tr_+ (\slashed{1} \slashed{2} \slashed{3} \slashed{l_4} \slashed{l_3} \slashed{n})}{(k_2 + l_1)^2 (l_2 + l_3)^2 (k_3 + l_4)^2 (k_n + l_3)^2},
\end{equation} where $\tr_+ (\slashed{a} \cdots) = \tfrac 1 2 \tr((1 + \gamma_5) \gamma^\mu\cdots) a_\mu \cdots$.  This can be further expanded and rearranged to give the following remarkably simple result\footnote{In writing the result below we have dropped all the parity odd contributions.  This is equivalent to the replacement $\tr_+ (\cdots) \to \tfrac 1 2 \tr (\cdots)$.}
\begin{equation}
  \label{eq:result-cut-a}
  \frac {s_{1 2}} 2 \Biggl[s_{1 2} s_{n 1} \raisebox{-5ex}{\includegraphics[height=10ex]{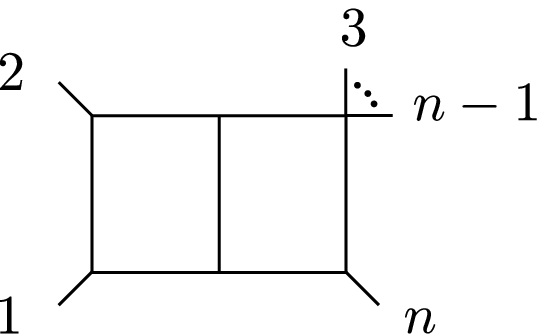}} +
 s_{1 2} s_{2 3} \raisebox{-6ex}{\includegraphics[height=10ex]{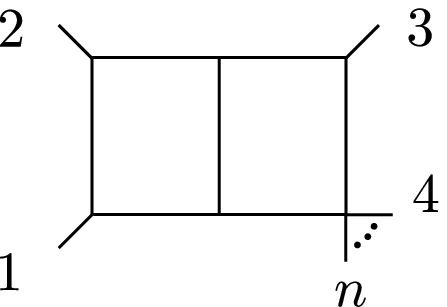}} +
 (s_{1 2 3} s_{n 1 2} - s_{1 2} s_{n 1 2 3}) \raisebox{-7ex}{\includegraphics[height=14ex]{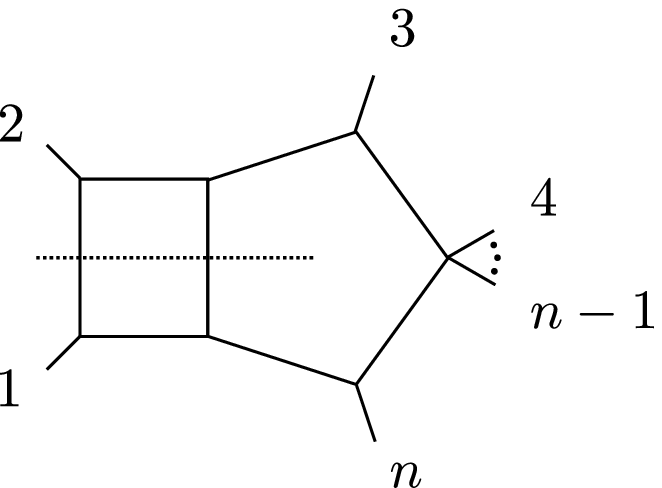}}\Biggr].
\end{equation}

The cut (b) in Fig.~\ref{fig:npt-cut} can be computed similarly.  After dividing by the tree, rationalizing the denominators and forming a trace from the spinor products, the result for the cut (b) becomes
\begin{equation}
  \label{eq:trace-cut-b}
  -s_{1 \cdots r} \frac {\tr_+(\slashed{l_1} \slashed{r} \slashed{(r+1)} \slashed{l_4} \slashed{l_3} \slashed{n} \slashed{1} \slashed{l_2})}{(k_1 + l_2)^2 (k_r + l_1)^2 (l_2 + l_3)^2 (k_{r+1} + l_4)^2 (k_n + l_3)^2}.
\end{equation}  This looks very similar to \cite[eq.~3.7]{Bern:2008ap}, and can be computed in a similar manner.  Again, dropping the odd part contribution and rewriting everything in dual variable language, we get

{\allowdisplaybreaks
\begin{equation}
  \begin{aligned}
    -\frac {x_{1,r+1}^2} 4 \Biggl\{& -\frac{x_{1,r+1}^2 x_{2 n}^2}{x_{2 p}^2 x_{n q}^2 x_{p q}^2} +\frac{x_{1 r}^2 x_{r+1 n}^2 - x_{1,r+1}^2 x_{r n}^2}{x_{n q}^2 x_{p q}^2 x_{r p}^2} + \frac{x_{1,r+2}^2 x_{2,r+1}^2-x_{1,r+1}^2 x_{2,r+2}^2}{x_{2 p}^2 x_{p q}^2 x_{r+2,q}^2} -\frac{x_{1,r+1}^2 x_{r,r+2}^2}{x_{p q}^2 x_{r p}^2 x_{r+2,q}^2} +\\
   & + \frac{\left(x_{1,r+1}^2 x_{2 r}^2-x_{1 r}^2 x_{2,r+1}^2\right)
   x_{n p}^2}{x_{2 p}^2 x_{n q}^2 x_{p q}^2 x_{r p}^2} + \frac 1 {x_{2 p}^2 x_{n q}^2 x_{r p}^2 x_{r+2,q}^2} \Bigl(-x_{1,r+2}^2 x_{2,r+1}^2 x_{r n}^2-x_{1,r+1}^2 x_{2,r+2}^2
   x_{r n}^2-\\&-x_{1,r+1}^2 x_{2 n}^2 x_{r,r+2}^2+x_{1,r+2}^2 x_{2 r}^2
   x_{r+1,n}^2-x_{1 r}^2 x_{2,r+2}^2 x_{r+1,n}^2-x_{1,r+1}^2 x_{2 r}^2
   x_{r+2,n}^2-x_{1 r}^2 x_{2,r+1}^2 x_{r+2,n}^2\Bigr) +\\
  & + \frac{x_{2 q}^2 \left(x_{1,r+1}^2 x_{r+2,n}^2-x_{1,r+2}^2
   x_{r+1,n}^2\right)}{x_{2 p}^2 x_{n q}^2 x_{p q}^2 x_{r+2,q}^2} + \frac{x_{r q}^2 \left(x_{1,r+1}^2 x_{r+2,n}^2-x_{1,r+2}^2
   x_{r+1,n}^2\right)}{x_{n q}^2 x_{p q}^2 x_{r p}^2 x_{r+2,q}^2} +\\&+ \frac{\left(x_{1,r+1}^2 x_{2 r}^2-x_{1 r}^2 x_{2,r+1}^2\right)
   x_{r+2,p}^2}{x_{2 p}^2 x_{p q}^2 x_{r p}^2 x_{r+2,q}^2} + \frac 1 {x_{2 p}^2 x_{n q}^2 x_{p q}^2 x_{r p}^2 x_{r+2,q}^2} \Bigl(\bigl(x_{1,r+2}^2 x_{2,r+1}^2 -x_{1,r+1}^2 x_{2,r+2}^2\bigr) x_{n p}^2 x_{r q}^2-\\&-x_{1,r+1}^2 x_{2 n}^2 x_{r+2,p}^2
   x_{r q}^2-x_{1,r+1}^2 x_{2 q}^2 x_{n p}^2 x_{r,r+2}^2+(x_{1 r}^2 x_{r+1,n}^2 -x_{1,r+1}^2 x_{r n}^2) x_{r+2,p}^2 x_{2 q}^2\Bigr)\Biggr\}.
  \end{aligned}
\end{equation}}

The even part of this cut can be expressed in terms of dual conformal integrals as well.  In fact, as can be seen from the formulas above, only a very restricted set of dual conformal integrals appear.

Before concluding this section let us note that results obtained above for the all-$n$ cuts are in agreement with both the results in Sec.~\ref{sec:seven-point} and in ref.~\cite{Bern:2008ap}.

\section{A leg addition rule}
\label{sec:leg-add}

Some heuristic rules like the rung rule (see ref.~\cite{Bern:1997nh}) and the box substitution rule (see ref.~\cite{Bern:2007ct}) enable one to get some integrals in the $L+1$-loop amplitude starting with the $L$-loop integrals.  Below we propose the leg addition rule which enables us to get some $n+1$-point integrals in the MHV amplitude starting with some $n$-point integrals.

By looking at eqs.~\eqref{eq:trace-cut-a}, \eqref{eq:trace-cut-b}, we observe that a large number of external lines don't appear at all in the formulas (they cancel in the ratio to the tree amplitude).  For example, the eq.~\eqref{eq:trace-cut-a} does not depend on the external momenta $k_4, \dotsc, k_{n-1}$, but only on the sum $k_4 + \dotso + k_{n-1}$ which is linked by momentum conservation to the sum $k_n + k_1 + k_2 + k_3$.  A similar statement holds for the eq.~\eqref{eq:trace-cut-b}.

In particular, the eq.~\eqref{eq:trace-cut-a} does not depend on the number of external lines inserted between the legs carrying momenta $k_3$ and $k_n$ (if the inserted leg does not alter the numbering of momenta).  In dual variable language, this means that the cuts in Fig.~\ref{fig:npt-cut} only depend on a small set of dual variables.

\begin{figure}
  \centering
  \includegraphics[width=\textwidth]{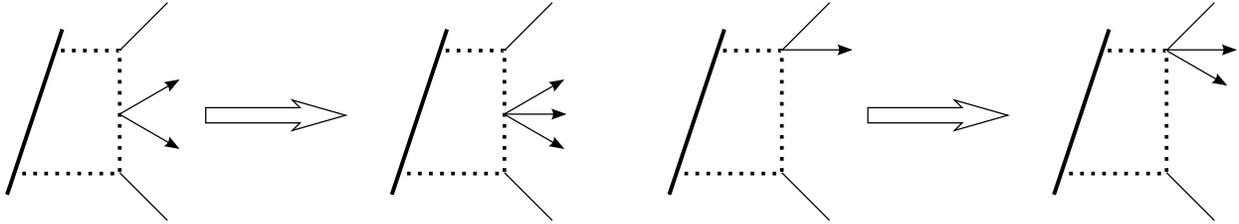}
  \caption{Two examples of applying the leg addition rule.  The dotted line is an internal line, carrying loop momentum.  When computing the coefficients, the sum of momenta marked by arrows before the transformation must be replaced with the sum of momenta after the transformation.}
  \label{fig:leg-addition}
\end{figure}

Based on the observation that the cuts depend on some of the momenta only through some linear combinations, it is easy to see that one can even change the number of momenta without changing the cut (provided that momentum is conserved, of course).  The leg addition rule is inspired by this observation.  The way this works in practice is best illustrated by a picture (see Fig.~\ref{fig:leg-addition}).\footnote{It is customary to write the amplitude in terms of sums of integrals times integral coefficients over cyclic and anti-cyclic permutations of the external legs.  Sometimes it is necessary to include symmetry factors to avoid multiple counting.  The leg addition rule formulated above can break or enhance the symmetry of the integral we start with and that has to be taken into account when writing the full amplitude.}

\begin{figure}
  \centering
  \includegraphics[width=\textwidth]{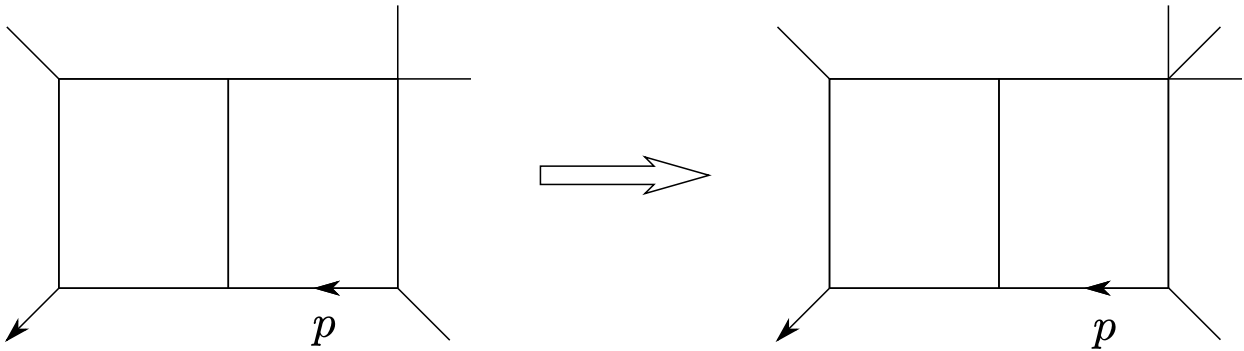}
  \caption{An example of relating the coefficients of two integrals by     applying the leg addition rule.  The external legs are numbered clockwise with $k_1$ indicated by an arrow.  All the external momenta are taken to be outgoing.}
  \label{fig:leg-addition-ex}
\end{figure}

Let us give an example to test that the leg addition rule works.  We choose an example where the odd parts are known analytically, to show that this also works for the odd part (see Fig.~\ref{fig:leg-addition-ex}).

The coefficient of the five-point integral in Fig.~\ref{fig:leg-addition-ex}, in the form it was presented\footnote{Here we have multiplied by four the result presented in ref.~\cite{Cachazo:2008vp}.} in ref.~\cite{Cachazo:2008vp}, is
\begin{equation}
  \label{eq:ls-5-pt}
  C_{\text{$5$-pt}} = \frac {4 s_{1 2}^2 s_{5 1} (p^{(1)} + k_{4 5})^2}{(p^{(1)} + k_{4 5})^2 - (p^{(2)} + k_{4 5})^2},
\end{equation} while the coefficient of the six-point integral, in the form it was presented in ref.~\cite{Cachazo:2008hp} is given by
\begin{equation}
  \label{eq:ls-6-pt}
  C_{\text{$6$-pt}} = \frac {4 s_{1 2}^2 s_{6 1} (p^{(1)} + k_{4 5 6})^2}{(p^{(1)} + k_{4 5 6})^2 - (p^{(2)} + k_{4 5 6})^2}.
\end{equation}  In the formulas above $p^{(1)}$ and $p^{(2)}$ are solutions to the leading singularity equations.  It is easy to see that the coefficient in eq.~\eqref{eq:ls-5-pt} is transformed to the coefficient in eq.~\eqref{eq:ls-6-pt} by the replacements $k_4 \to k_4 + k_5$ and $k_5 \to k_6$ (where the last transformation is due to the relabeling of momenta).

\section{Comments and discussion}

In conclusion, we computed the even part of two-loop seven-point planar MHV amplitude.  We also computed two all-$n$ cuts, as a modest step towards finding the full two-loop MHV amplitude of the $\mathcal{N}=4$ theory.  We found that, out of all the possible dual conformal integrals contributing to the $n$-point amplitude, only a very restricted set seems to appear.

Based on the form of the MHV amplitude, we proposed a ``leg addition rule'' as a counterpart to the well known rung rule and box substitution rule.

This work might serve as a starting point in understanding the correspondence between the integrals appearing in the evaluation of the MHV amplitude and the integrals appearing in the evaluation of the Wilson loop.  The master integrals appearing in the Wilson loop computation have been listed and computed numerically in ref.~\cite{Anastasiou:2009kn}.  Note that the integrals appearing in the Wilson loop computations are much easier to evaluate than the ones appearing in the scattering amplitude.  Therefore, establishing a correspondence between the two sets of integrals could have important practical consequences for the evaluation of amplitudes.

We are left with a number of open questions.  One pressing question is what is the basis of integrals for the two-loop MHV amplitudes.  This question has received a partial answer in this paper.  We hope to revisit this question in the future.

Another question concerns the odd parts and the $\mu$ terms.  Some odd parts have been efficiently computed so far (see refs.~\cite{Cachazo:2008vp, Cachazo:2008hp, Spradlin:2008uu}), but at six points (and probably at higher points as well) one encounters the thorny problem of linear dependence of the integrals.\footnote{For the even part we have a preferred basis of integrals containing the dual conformal integrals.  In all known examples, and also in the new cases analyzed in this paper, there seem to be no reduction identities relating only dual conformal integrals (i.e.\ the coefficients of the dual conformal integrals are determined uniquely).  The situation is less favorable for the odd parts of the amplitudes, because in that case we don't know of any canonical choice for the basis of integrals.  In this case the integral coefficients are not uniquely determined and they depend on the choice of which integrals to eliminate from the ``basis'' by using reduction identities (see ref.~\cite{Cachazo:2008hp} for an example of this situation).}  In this regard, it would be interesting to see if the leg addition rule proposed in Sec.~\ref{sec:leg-add} can be used to constrain the choice of basis for the odd part.

So far, the discussion was restricted to MHV amplitudes.  But one would also like to understand the NMHV amplitudes.  Related to this, one would like to understand how to ``dress'' the Wilson loop with helicity information.

\section*{Acknowledgments}

I am grateful to Marcus Spradlin for collaboration in the initial stages of this project, for his insight and for his joviality.  If it weren't for his contagious optimism, this paper might have never appeared.  I am also grateful to Zvi Bern, Lance Dixon, David Kosower, Radu Roiban and Anastasia Volovich for previous collaboration on related topics.  I wish to thank Marcus Spradlin and Anastasia Volovich for reading a copy of the draft and making constructive comments.

This research is supported in part by the US Department of Energy under contract DE-FG02-91ER40688 and by the US National Science Foundation under grant PHY-0643150.


\begin{thebibliography}{99}
\bibitem{'tHooft:1973jz}
  G.~'t Hooft,
  ``A PLANAR DIAGRAM THEORY FOR STRONG INTERACTIONS,''
  Nucl.\ Phys.\  B {\bf 72}, 461 (1974).

\bibitem{Maldacena:1997re}
  J.~M.~Maldacena,
  ``The large N limit of superconformal field theories and supergravity,''
  Adv.\ Theor.\ Math.\ Phys.\  {\bf 2} (1998) 231
  [Int.\ J.\ Theor.\ Phys.\  {\bf 38} (1999) 1113]
  [arXiv:hep-th/9711200].
\bibitem{Gubser:1998bc}
  S.~S.~Gubser, I.~R.~Klebanov and A.~M.~Polyakov,
  ``Gauge theory correlators from non-critical string theory,''
  Phys.\ Lett.\  B {\bf 428}, 105 (1998)
  [arXiv:hep-th/9802109].
\bibitem{Witten:1998qj}
  E.~Witten,
  ``Anti-de Sitter space and holography,''
  Adv.\ Theor.\ Math.\ Phys.\  {\bf 2}, 253 (1998)
  [arXiv:hep-th/9802150].

\bibitem{Minahan:2002ve}
  J.~A.~Minahan and K.~Zarembo,
  ``The Bethe-ansatz for N = 4 super Yang-Mills,''
  JHEP {\bf 0303}, 013 (2003)
  [arXiv:hep-th/0212208].

\bibitem{Bena:2003wd}
  I.~Bena, J.~Polchinski and R.~Roiban,
  ``Hidden symmetries of the AdS(5) x S**5 superstring,''
  Phys.\ Rev.\  D {\bf 69}, 046002 (2004)
  [arXiv:hep-th/0305116].

\bibitem{Witten:2003nn}
  E.~Witten,
  ``Perturbative gauge theory as a string theory in twistor space,''
  Commun.\ Math.\ Phys.\  {\bf 252}, 189 (2004)
  [arXiv:hep-th/0312171].

\bibitem{Cachazo:2004kj}
  F.~Cachazo, P.~Svrcek and E.~Witten,
  ``MHV vertices and tree amplitudes in gauge theory,''
  JHEP {\bf 0409}, 006 (2004)
  [arXiv:hep-th/0403047].

\bibitem{Britto:2005fq}
  R.~Britto, F.~Cachazo, B.~Feng and E.~Witten,
  ``Direct Proof Of Tree-Level Recursion Relation In Yang-Mills Theory,''
  Phys.\ Rev.\ Lett.\  {\bf 94}, 181602 (2005)
  [arXiv:hep-th/0501052].

\bibitem{ArkaniHamed:2009si}
  N.~Arkani-Hamed, F.~Cachazo, C.~Cheung and J.~Kaplan,
  ``The S-Matrix in Twistor Space,''
  arXiv:0903.2110 [hep-th].

\bibitem{Mason:2009sa}
  L.~Mason and D.~Skinner,
  ``Scattering Amplitudes and BCFW Recursion in Twistor Space,''
  arXiv:0903.2083 [hep-th].

\bibitem{Drummond:2009fd}
  J.~M.~Drummond, J.~M.~Henn and J.~Plefka,
  ``Yangian symmetry of scattering amplitudes in N=4 super Yang-Mills theory,''
  arXiv:0902.2987 [hep-th].

\bibitem{Anastasiou:2003kj}
  C.~Anastasiou, Z.~Bern, L.~J.~Dixon and D.~A.~Kosower,
  ``Planar amplitudes in maximally supersymmetric Yang-Mills theory,''
  Phys.\ Rev.\ Lett.\  {\bf 91} (2003) 251602
  [arXiv:hep-th/0309040].

\bibitem{Bern:2005iz}
  Z.~Bern, L.~J.~Dixon and V.~A.~Smirnov,
  ``Iteration of planar amplitudes in maximally supersymmetric Yang-Mills
  theory at three loops and beyond,''
  Phys.\ Rev.\  D {\bf 72} (2005) 085001
  [arXiv:hep-th/0505205].

\bibitem{Bern:2008ap}
  Z.~Bern, L.~J.~Dixon, D.~A.~Kosower, R.~Roiban, M.~Spradlin, C.~Vergu and A.~Volovich,
  ``The Two-Loop Six-Gluon MHV Amplitude in Maximally Supersymmetric Yang-Mills
  Theory,''
  Phys.\ Rev.\  D {\bf 78}, 045007 (2008)
  [arXiv:0803.1465 [hep-th]].

\bibitem{Bern:2006vw}
  Z.~Bern, M.~Czakon, D.~A.~Kosower, R.~Roiban and V.~A.~Smirnov,
  ``Two-loop iteration of five-point N = 4 super-Yang-Mills amplitudes,''
  Phys.\ Rev.\ Lett.\  {\bf 97}, 181601 (2006)
  [arXiv:hep-th/0604074].

\bibitem{Bern:2007ct}
  Z.~Bern, J.~J.~M.~Carrasco, H.~Johansson and D.~A.~Kosower,
  ``Maximally supersymmetric planar Yang-Mills amplitudes at five loops,''
  Phys.\ Rev.\  D {\bf 76}, 125020 (2007)
  [arXiv:0705.1864 [hep-th]].

\bibitem{Bern:1994zx}
  Z.~Bern, L.~J.~Dixon, D.~C.~Dunbar and D.~A.~Kosower,
  ``One-Loop n-Point Gauge Theory Amplitudes, Unitarity and Collinear Limits,''
  Nucl.\ Phys.\  B {\bf 425}, 217 (1994)
  [arXiv:hep-ph/9403226].

\bibitem{Bern:1997sc}
  Z.~Bern, L.~J.~Dixon and D.~A.~Kosower,
  ``One-loop amplitudes for e+ e- to four partons,''
  Nucl.\ Phys.\  B {\bf 513}, 3 (1998)
  [arXiv:hep-ph/9708239].

\bibitem{Bern:1994cg}
  Z.~Bern, L.~J.~Dixon, D.~C.~Dunbar and D.~A.~Kosower,
  ``Fusing gauge theory tree amplitudes into loop amplitudes,''
  Nucl.\ Phys.\  B {\bf 435}, 59 (1995)
  [arXiv:hep-ph/9409265].

\bibitem{Bern:1997nh}
  Z.~Bern, J.~S.~Rozowsky and B.~Yan,
  ``Two-loop four-gluon amplitudes in N = 4 super-Yang-Mills,''
  Phys.\ Lett.\  B {\bf 401}, 273 (1997)
  [arXiv:hep-ph/9702424].

\bibitem{Alday:2007hr}
  L.~F.~Alday and J.~M.~Maldacena,
  ``Gluon scattering amplitudes at strong coupling,''
  JHEP {\bf 0706} (2007) 064
  [arXiv:0705.0303 [hep-th]].

\bibitem{Alday:2007he}
  L.~F.~Alday and J.~Maldacena,
  ``Comments on gluon scattering amplitudes via AdS/CFT,''
  JHEP {\bf 0711} (2007) 068
  [arXiv:0710.1060 [hep-th]].

\bibitem{Drummond:2006rz}
  J.~M.~Drummond, J.~Henn, V.~A.~Smirnov and E.~Sokatchev,
  ``Magic identities for conformal four-point integrals,''
  JHEP {\bf 0701}, 064 (2007)
  [arXiv:hep-th/0607160].

\bibitem{Drummond:2007aua}
  J.~M.~Drummond, G.~P.~Korchemsky and E.~Sokatchev,
  ``Conformal properties of four-gluon planar amplitudes and Wilson loops,''
  Nucl.\ Phys.\  B {\bf 795}, 385 (2008)
  [arXiv:0707.0243 [hep-th]].

\bibitem{Drummond:2007cf}
  J.~M.~Drummond, J.~Henn, G.~P.~Korchemsky and E.~Sokatchev,
  ``On planar gluon amplitudes/Wilson loops duality,''
  Nucl.\ Phys.\  B {\bf 795}, 52 (2008)
  [arXiv:0709.2368 [hep-th]].

\bibitem{Drummond:2007au}
  J.~M.~Drummond, J.~Henn, G.~P.~Korchemsky and E.~Sokatchev,
  ``Conformal Ward identities for Wilson loops and a test of the duality with
  arXiv:0712.1223 [hep-th].

\bibitem{Drummond:2007bm}
  J.~M.~Drummond, J.~Henn, G.~P.~Korchemsky and E.~Sokatchev,
  ``The hexagon Wilson loop and the BDS ansatz for the six-gluon amplitude,''
  Phys.\ Lett.\  B {\bf 662}, 456 (2008)
  [arXiv:0712.4138 [hep-th]].

\bibitem{Drummond:2008aq}
  J.~M.~Drummond, J.~Henn, G.~P.~Korchemsky and E.~Sokatchev,
  ``Hexagon Wilson loop = six-gluon MHV amplitude,''
  arXiv:0803.1466 [hep-th].

\bibitem{Drummond:2008vq}
  J.~M.~Drummond, J.~Henn, G.~P.~Korchemsky and E.~Sokatchev,
  ``Dual superconformal symmetry of scattering amplitudes in N=4
  super-Yang-Mills theory,''
  arXiv:0807.1095 [hep-th].

\bibitem{Drummond:2008bq}
  J.~M.~Drummond, J.~Henn, G.~P.~Korchemsky and E.~Sokatchev,
  ``Generalized unitarity for N=4 super-amplitudes,''
  arXiv:0808.0491 [hep-th].

\bibitem{Drummond:2008cr}
  J.~M.~Drummond and J.~M.~Henn,
  ``All tree-level amplitudes in N=4 SYM,''
  arXiv:0808.2475 [hep-th].

\bibitem{Cachazo:2008hp}
  F.~Cachazo, M.~Spradlin and A.~Volovich,
  ``Leading Singularities of the Two-Loop Six-Particle MHV Amplitude,''
  Phys.\ Rev.\  D {\bf 78}, 105022 (2008)
  [arXiv:0805.4832 [hep-th]].

\bibitem{Cachazo:2008vp}
  F.~Cachazo,
  ``Sharpening The Leading Singularity,''
  arXiv:0803.1988 [hep-th].

\bibitem{Cachazo:2008dx}
  F.~Cachazo and D.~Skinner,
  ``On the structure of scattering amplitudes in N=4 super Yang-Mills and N=8
  supergravity,''
  arXiv:0801.4574 [hep-th].

\bibitem{Buchbinder:2005wp}
  E.~I.~Buchbinder and F.~Cachazo,
  ``Two-loop amplitudes of gluons and octa-cuts in N = 4 super Yang-Mills,''
  JHEP {\bf 0511}, 036 (2005)
  [arXiv:hep-th/0506126].

\bibitem{ArkaniHamed:2008gz}
  N.~Arkani-Hamed, F.~Cachazo and J.~Kaplan,
  ``What is the Simplest Quantum Field Theory?,''
  arXiv:0808.1446 [hep-th].

\bibitem{Cachazo:2006tj}
  F.~Cachazo, M.~Spradlin and A.~Volovich,
  ``Iterative structure within the five-particle two-loop amplitude,''
  Phys.\ Rev.\  D {\bf 74}, 045020 (2006)
  [arXiv:hep-th/0602228].

\bibitem{Spradlin:2008uu}
  M.~Spradlin, A.~Volovich and C.~Wen,
  ``Three-Loop Leading Singularities and BDS Ansatz for Five Particles,''
  Phys.\ Rev.\  D {\bf 78}, 085025 (2008)
  [arXiv:0808.1054 [hep-th]].

\bibitem{Anastasiou:2009kn}
  C.~Anastasiou, A.~Brandhuber, P.~Heslop, V.~V.~Khoze, B.~Spence and G.~Travaglini,
  ``Two-Loop Polygon Wilson Loops in N=4 SYM,''
  arXiv:0902.2245 [hep-th].

\bibitem{Brandhuber:2007yx}
  A.~Brandhuber, P.~Heslop and G.~Travaglini,
  ``MHV Amplitudes in N=4 Super Yang-Mills and Wilson Loops,''
  Nucl.\ Phys.\  B {\bf 794}, 231 (2008)
  [arXiv:0707.1153 [hep-th]].

\bibitem{unpublished:2008}
  C.~Vergu,
  unpublished

\bibitem{Bern:2008qj}
  Z.~Bern, J.~J.~M.~Carrasco and H.~Johansson,
  Phys.\ Rev.\  D {\bf 78}, 085011 (2008)
  [arXiv:0805.3993 [hep-ph]].

\bibitem{Ricci:2007eq}
  R.~Ricci, A.~A.~Tseytlin and M.~Wolf,
  JHEP {\bf 0712}, 082 (2007)
  [arXiv:0711.0707 [hep-th]].

\bibitem{Berkovits:2008ic}
  N.~Berkovits and J.~Maldacena,
  ``Fermionic T-Duality, Dual Superconformal Symmetry, and the Amplitude/Wilson
  Loop Connection,''
  JHEP {\bf 0809}, 062 (2008)
  [arXiv:0807.3196 [hep-th]].

\bibitem{Beisert:2008iq}
  N.~Beisert, R.~Ricci, A.~A.~Tseytlin and M.~Wolf,
  ``Dual Superconformal Symmetry from AdS5 x S5 Superstring Integrability,''
  Phys.\ Rev.\  D {\bf 78}, 126004 (2008)
  [arXiv:0807.3228 [hep-th]].

\bibitem{Dixon:2008tu}
  L.~J.~Dixon,
  ``Gluon scattering in N=4 super-Yang-Mills theory from weak to strong
  coupling,''
  PoS {\bf RADCOR2007}, 056 (2007)
  [arXiv:0803.2475 [hep-th]].

\bibitem{Alday:2008yw}
  L.~F.~Alday and R.~Roiban,
  ``Scattering Amplitudes, Wilson Loops and the String/Gauge Theory
  Correspondence,''
  Phys.\ Rept.\  {\bf 468}, 153 (2008)
  [arXiv:0807.1889 [hep-th]].

\bibitem{Bern:2007dw}
  Z.~Bern, L.~J.~Dixon and D.~A.~Kosower,
  ``On-Shell Methods in Perturbative QCD,''
  Annals Phys.\  {\bf 322}, 1587 (2007)
  [arXiv:0704.2798 [hep-ph]].

\bibitem{Aharony:2008ug}
  O.~Aharony, O.~Bergman, D.~L.~Jafferis and J.~Maldacena,
  JHEP {\bf 0810}, 091 (2008)
  [arXiv:0806.1218 [hep-th]].

\bibitem{Agarwal:2008pu}
  A.~Agarwal, N.~Beisert and T.~McLoughlin,
  arXiv:0812.3367 [hep-th].

\end{thebibliography}
\end{document}